\documentstyle[titlepage,12pt]{article}
\begin{document}
\title {Exponential - Potential Scalar Field Universes I. The Bianchi I Models
}  \author {
J. M. Aguirregabiria, A. Feinstein and J. Ib\'a\~nez\\ Dept.  de F\'{\i}sica
Te\'orica,
Universidad del Pa\'{\i}s Vasco, Bilbao, Spain.} \maketitle

\begin {abstract}

We obtain a general exact solution of the Einstein field equations for
the anisotropic Bianchi type I universes filled with an exponential-potential
scalar field  and study their dynamics. It is shown, in agreement with previous
studies,  that for a wide range of initial conditions the late-time behaviour
of
the models is that of a  power-law inflating FRW universe. This property,
  does not hold, in contrast, when some degree of inhomogeneity is
introduced, as discussed in our following paper II.

\end {abstract}

\section{Introduction}

Two of the present authors have recently studied exact model universes filled
with an exponential-potential scalar field and have found \cite{1}, \cite{2} a
persistent non-inflationary behaviour in some exact  Bianchi types III and VI,
as
well as in some  cosmological models which may be thought as inhomogeneous
generalizations of Bianchi type I cosmologies. These studies cast a doubt on
the
so-called  ``inflationary paradigm" which,  while never properly formulated,
vaguely states that  the universe had undergone a period of inflationary
expansion which is not only a must to solve a host of problems of
standard cosmology, but is rather a ``typical" dynamical behaviour common to  a
wide class of scalar field cosmologies.

One of the main reasons the inflationary universe scenario is considered to be
attractive \cite{3} is
that it has brought back to life Misner's hope \cite{4} to explain the
large-scale homogeneity,  implied by the measurements of the cosmic mirowave
background radiation,  without a need to impose  very special conditions on the
initial expansion of the universe. Strangely enough, however,  most of the work
on inflation is done in the framework of isotropic and homogeneous FRW
universes
(Olive \cite{5} and  the references there in).

This paper deals with {\em exact solutions} of the Einstein field equations. We
show
how one may obtain a general exact solution for Bianchi type I  cosmologies
with
an exponential-potential scalar field. Apart that one obviously needs a
specific
model for the potential to solve exactly the Einstein   equations,  it is
convenient to concentrate on this sort of the potential,  often used in
inflationary analysis,  for the following main reasons:

1. Exponential-potential scalar field introduces rather a  small additional
degree of nonlinearity into Einstein field equations, as it will
be explained later, so that possibilities exist to solve these equations
analytically in a variety of cases. In the case of  cosmological models with
Bianchi type I  spatial symmetry one can obtain a general solution.

2. It seems that the exponential potentials for scalar fields  arise
  \cite{6}
in many theories such as Jordan-Brans-Dicke theory, the superstring theory,
Salam-Sezgin theory  and others.

This work represents an introductory step towards our following paper II  where
 the homogeneity will be broken in one direction to study the
inhomogeneity effects on the late-time behaviour of scalar field cosmological
models.

The main result of this paper is to confirm, on the basis of {\em exact
solutions} we obtain, previous
 qualitative and numerical studies by other authors \cite{7} and  the pattern
of the late-time behaviour of the exponential-potential scalar field
cosmologies. This agreement is only true, however, as long as the underlying
geometry of the cosmological model is as simple as that of Bianchi type I. When
the  spatial symmetry group is more complicated, as for example in Bianchi III
and Bianchi VI models as well as when the spatial homogeneity is broken the
dynamics of the models is different. This, however,  will be
discussed in a separate paper II.

In the following Section II we discuss the  geometry and the matter content of
the models. In Section III the Einstein equations are considered and solved.
Some representative solutions are given in the section IV, while in the Section
V the late-time behaviour of the generic solutions is studied qualitatively
and numerically. In Section VI we conclude and summarise our results.

\section{The Bianchi I Exponential-Potential Scalar Field Cosmologies}

The usual synchronous form for a general Bianchi type I element is given by
\begin{equation}
ds^2 = -dT^2 + a_1^2(T)\, dx^2 + a_2^2(T)\, dy^2 + a_3^2(T)\, dz^2 \,,
\end {equation}
representing the anisotropic generalization of the spatially flat FRW universe
expanding
differently in \(x, y\), and \(z\) directions. While the form (1) is frequently
used to study the dynamical behaviour of the cosmological model, we have found
it
convenient, in order to obtain exact analytic solutions, to cast the
metric into a following, so-to-say, semi-conformal form
\begin{equation}
ds^2 = e^{f(t)} (-dt^2 + dz^2) + G(t)\, ( e^{p(t)} dx^2 +  e^{-p(t)} dy^2) \,.
\end{equation}
This form of the metric is of particular interest when studying the
inhomogeneous generalizations of
the line element (1), and is obtained from the Eq.(1) by the following
identification
\begin{equation}
dt = \frac{dT}{a_3}\, ,  e^{f}= a_3^2\, ,  G = a_1 a_2\, ,  e^{p} =
\frac{a_1}{a_2}\,.
\end{equation}

Since from now on we will be working with the line element (2) exclusively it
is worthwhile to
specify some cases of interest corresponding to particular values the metric
functions \(f\) ,
\(G \) and \(p\) can take.

When the function \(p\) vanishes globally the spacetime expands equally in axes
\(x\)  and \(y\) in which case the spacetime is called  Locally Rotationally
Symmetric (LRS). If in addition to \(p\) = 0 one has  \(G= e^{f(t)}\) then the
metric is of the FRW form. In fact the only way the metric (2) may approach the
FRW solution is when  \(e^{f(t)} \sim G\) together with \(p\sim 0\).

The line element (2) admits important vacuum solutions
\begin{equation}
 G = t\, ,\quad  p = \kappa\log t\,  ,\quad  f= \frac{\kappa^2-1}{2}\log t\,,
\end{equation}
which are of Kasner type and always anisotropic apart from two ``degenerate
cases" \(\kappa=\pm 1\) which are disguised Minkowski line elements.

The energy-momentum tensor for the scalar field driven by the potential
\(V(\phi)\) is given by
\begin{equation}
T_{\alpha\beta}=\phi_{,\alpha}\phi_{,\beta}-g_{\alpha\beta}\,
\left( \frac{1}{2}\phi_{,\gamma}\phi^{,\gamma}+V(\phi)\right)\,,
\end{equation}
and as long as the attention is concentrated on the homogeneous space-times the
Eq.(5) may be
rewritten in the perfect fluid form
\begin{equation}
T_{\alpha\beta}= (p+\rho) u_{\alpha}u_{\beta}+pg_{\alpha\beta}\,,
\end{equation}
where
\begin{equation}
u_\alpha=\frac{\phi_{,\alpha}}{\sqrt{-\phi_{,\gamma}\phi^{,\gamma}}}\,.
\end{equation}
together with
\begin{eqnarray}
\rho & = &  -\frac{1}{2} \phi_{,\gamma}\phi^{,\gamma} + V(\phi)\nonumber \\
  p & = &  -\frac{1}{2} \phi_{,\gamma}\phi^{,\gamma} - V(\phi)
\end{eqnarray}

The perfect fluid interpretation of the scalar field, while not obligatory, is
very useful to study the kinematical behaviour of the cosmological models. It
is
convenient to introduce \cite{8} the expansion
\begin{equation}
\Theta = u_{\mu ; \nu} g^{\mu \nu}
\end{equation}
and the deceleration parameter
\begin{equation}
q = -3 \Theta^2\, \left(\Theta_{; \alpha} u^{\alpha}
+\frac{1}{3}\Theta^2\right).
\end{equation}

The sign of the deceleration parameter indicates as to whether the
cosmological
model inflates. The positive sign corresponds to ``standard"  decelerating
models
whereas the negative sign indicates inflation.

For the models to isotropize at late times the shear measured with respect to
expansion rate must  vanish asymptotically  \cite{9}.  The models
defined by the line element (2) are simple enough so that one may get a precise
idea
as to whether the spacetime  isotropizes just from looking at the asymptotic
behaviour  of the line element.

\section{Solving  the Einstein Equations }

The Einstein Equations for the metric given by the line element (2) and the
matter
specified by the  stress-energy tensor (5) are given by
\begin{equation}
\ddot{\phi}+ \frac{\dot{G}}{G}\,\dot{\phi} +e^{f}\,
\frac{\partial{V}}{\partial{\phi}} = 0 \,,
 \end{equation}
\begin{equation}
\frac{\ddot G}{G} = 2\, e^{f}\, V\,,
\end{equation}
\begin{equation}
\ddot{p} + \frac{\dot G}{G}\, \dot{p} = 0 \,,
\end{equation}
\begin{equation}
\frac{\ddot G}{G} - \frac{1}{2}\left(\frac{\dot G}{G}\right)^2 - \frac{\dot
G}{G}
\,\dot{f} + \frac{1}{2}\dot{p}^2 = - \dot{\phi}^2\,,
\end{equation}
where  the Eq.~(11) is the Klein-Gordon equation for the scalar field.

Note that the Eq.~(13) is somewhat decoupled from other equations in the
sense that it is the same as in the vacuum case. The matter field acts
indirectly on this equation through the ``transitivity area" function \(G\).
Yet, one may immediately integrate the equation in terms of the function \(G\):
\begin{equation}
\dot{p} = \frac{a}{G} \,,
\end{equation}
where  \(a\) is an arbitrary integration constant.

Now we substitute the Eq.~(15) into the Eq.~(14) and use the fact that on
differentiating the Eq.~(12) and using it again  one gets, for \(V\neq 0\),
\begin{equation}
\dot{f} = \frac{{\buildrel\ldots\over G}}{\ddot G}  - \frac{\dot G}{G} -
\frac{\dot V}{V}\,.
\end{equation}
We then are left with the two following equations
\begin{equation}
\ddot{\phi} + \frac{\dot G}{G}\, \dot{\phi} = -  \frac{1}{2}\,
\frac{\ddot{G}}{GV}\,\frac{\partial V}{\partial \phi}
\end{equation}
\begin{equation}
\frac{\ddot G}{G} +\frac{1}{2}\, \frac{\dot G^2 + a^2}{G^2} -
\frac{{\buildrel\ldots\over G}\,
\dot G}{G \,\ddot G} + \frac{\dot G}{G}\, \frac{\ddot V}{V} = - \dot\phi^2\, ,
\end{equation}
where the Eq.~(17) has been obtained by substituting the Eq.~(12) into the
Eq.~(11).

We now specify the potential
\begin{equation}
V(\phi) = \Lambda e^{k \phi}
\end{equation}
Note that the Eqs.~(17) and (18) are not valid for the case  \(V = 0\),
in this case, however, the original system of equations is much simpler and its
general solution is given by
\begin{equation}
\Lambda = 0 ,\:  G=t,\: p = \alpha\log t,\: \phi = \beta\log t,\:
 f= \frac{\alpha^2 + 2\beta^2 - 1}{2}\log t\,.
\end{equation}
When \(\alpha\) = \(0\) and \(\beta = \pm\sqrt{3}/2\) the above solution
describes an isotropic spatially flat FRW universe.

Returning now to the general case \(\Lambda \neq 0\), the Klein-Gordon
equation (17) takes the form
\begin{equation}
\ddot{\phi} + \frac{\dot G}{G}\, \dot{\phi} = - \frac{k}{2}\, \frac{\ddot
G}{G}\, .
\end{equation}

In analogy with the Eq.~(13) one readily obtains from the Eq.~(21)
\begin{equation}
\dot{\phi} =   \frac{m}{G}- \frac{k}{2}\, \frac{\dot{G}}{G}\,,
\end{equation}
where \(m\) is an arbitrary constant.

It is important to note that the exponential potential saves one much trouble
with
the non-linearity of the Klein-Gordon equation. This is not only true for the
homogeneous Bianchi type I models but for any homogeneous or inhomogeneous
cosmological spacetime which may be cast into the ``semi-conformal'' form (2)
with the metric functions depending on \(z\) coordinate as well (technically
these are called the spacetimes with two commuting orthogonal  spacelike
Killing
vectors and include Bianchi types I, III, V, VI  and their unidirectional
inhomogeneous generalizations).

Substituting  \(\dot{\phi}\) given by the Eq.~(22) and the form of the
potential
into the Eq.~(18) we are left with a single equation
\begin{equation}
G\ddot G ^2 - {\buildrel\ldots\over G} \dot G G + \left( \frac{1}{2} -
\frac{k^2}{4}\right)
\,\ddot G \dot G^2 + \left(m^2 + \frac{a^2}{2}\right)\, \ddot G  = 0\,.
\end{equation}

The substitution
\begin{equation}
\dot G = y(G) \,,
\end{equation}
reduces the degree of the last equation, and after redefining some of the
constants
we finally get
\begin{equation}
 y'' + \left( K - \frac{M}{y^2}\right)\, \frac{1}{G}\, y' =  0\,,
\end{equation}
where
\begin{equation}
K = \frac{k^2}{4} - \frac{1}{2} ,\quad M =   m^2 +\frac{a^2}{2}\,.
\end{equation}
and \(y'= dy/dG\) .

The Eq.~(25) has the following first integral
\begin{equation}
G y' + (K-1) y + \frac{M}{y} = B\,,
\end{equation}
where B is an arbitrary constant. Integrating the Eq.~(27) one finally
obtains
\begin{equation}
G = \left[ \frac{2 (K-1)y + B - \sqrt{\Delta}}{2 (K-1)y + B + \sqrt{\Delta}}
\right]^{B/2(K-1) \sqrt{\Delta}} \left\{ N[(K-1)y^2 + By +
M]\right\}^{-1/2(K-1)},
\end{equation}
where
\begin{equation}
\Delta = B^2 - 4(K - 1) M \,,
\end{equation}
and \(N\) is yet another arbitrary constant of integration. The Eq.~(28) as
it stands is valid for \(\Delta\geq 0\). For \(\Delta<0\) the integral takes a
different form.

This formally concludes the solution. For, once \(G(y)\) is given by the
 Eq.~(28)  one may in principle invert the expression to get the \(y(G)\),
then using  \( t= \int \frac{dG}{y(G)}\)  and inverting again to obtain
\(G(t)\). From  \(G(t)\) one easily reconstructs all the metric functions:
\(p(t)\) from the Eq.~(15), \(\phi(t)\)  from the Eq.~(22) and \(f(t)\) from
the
Eq.~(12) for example.

Before turning to analyse the general solution given by the Eq.~(28) we will
present some
 explicit particular cases of interest in the following Section IV to
illustrate  the
above mentioned procedure.

\section{Explicit Exact Solutions}

In this Section we obtain explicitly some exact solutions of interest and
briefly discuss their behaviour. We start with the simplest ones.

\subsection{The FRW Universes}

The homogeneous and isotropic universes are obtained if one specifies
\(M=0\) and \(B=0\) in the Eq.~(27). Note that \(M=0\) automatically means
\(a=0\) in the Eq.~(15) which in  turn excludes the transversal part of
the gravitational field \(p\) restricting the class of models to LRS
ones. If, moreover, the constant \(B=0\) one always finishes with an
isotropic solution.

If \(K=0\) (\(k^2=2\)) one obtains the so called ``coasting solution''
\cite{10}:
\begin{equation}
ds^2 = \frac{A^2}{2 \Lambda} e^{At}(-dt^2+dx^2+dy^2+dz^2)
\end{equation}
and
\begin{equation}
\phi = -\frac{k}{2} A t
\end{equation}
This universe expands linearly in synchronous coordinates and has zero
deceleration parameter.

For \(K\neq 0\) one gets from the Eq.~(28)
\begin{equation}
G=t^{\lambda},\quad \lambda = \frac{4}{k^2-2}
\end{equation}
Two different classes of solutions appear. If \(\lambda=1\) one has
\(\Lambda=0\) and the solution is that described by the Eq.~(20)
representing the massless minimally coupled scalar field FRW universe. If,
however, \(\lambda\neq 1\) one gets
\begin{equation}
G=t^{\lambda},\; e^f=\frac{\lambda (\lambda-1)}{2\Lambda} t^\lambda,\; \phi =
-\frac{k\lambda}{2} \log t\, ,
\end{equation}
and the metric in synchronous form is given by
\begin{equation}
ds^2=-dT^2+T^{4/k^2} (dx^2+dy^2+dz^2)
\end{equation}

The sign of the deceleration parameter for the models described by the Eq.~(34)
depends on the quantity \(k^2-2\) so that the model inflates for \(k^2<2\)
while
decelerates for \(k^2>2\).

\subsection{The LRS Models}

Assuming now \(M=0\) but \(B\neq 0\) and taking \(K=0\) to obtain an
analytic expression we get from the integral (28)
\begin{equation}
G=e^t + B,\; e^f=\frac{1}{2\Lambda} e^t,\; \phi=-\frac{k}{2}\log (e^t+B).
\end{equation}
The metric in the synchronous form is given by
\begin{equation}
ds^2=-dT^2+T^2dz^2+(T^2+B)(dx^2+dy^2).
\end{equation}
Note that at early times this LRS model is anisotropic while at
\(T\rightarrow \infty\) approaches the linearly expanding FRW
universe.

The sign of the deceleration parameter is given by
\begin{equation}
-B (e^t-\frac{B}{3})\, ,
\end{equation}
and if the constant \(B<0\) the model never inflates! Note, however, that
the range of the parameter \(B\) might be restricted by the sign of the
potential \(V\). In the solution (36) the constant \(B\) measures the
deviations from isotropy and as long as \(B\) is negative the model does not
inflate independently of how negligible the deviation from the isotropy is.

In the case \(B>0\) the cosmological model decelerates until
\(t<\log\frac{B}{3}\) but inflates for later times. Another interesting
observation for this model is that they are non singular for \(B>0\). This
does not contradict the singularity theorems for, if \(B>0\) one has
\begin{equation}
\rho+3p=-\frac{4B}{(B+T^2)^2}<0\, ,
\end{equation}
and the energy condition is broken.

\subsection{Non Symmetric Solutions}

We put \(M\neq 0, B=0, N=-1/M\) as well as specify \(K=0\) again to obtain an
analytic expression. With these constants the integral (28) gives
\begin{equation}
G=\sinh \omega t,\quad \omega=\sqrt{M}.
\end{equation}
The transversal and longitudinal degrees of freedom are then given by
\begin{equation}
p=\frac{a}{\omega} \log \left(\tanh \frac{\omega t}{2}\right)
\end{equation}
and
\begin{equation}
f=\log (\sinh \omega t)-\frac{km}{\omega}\log \left(\tanh \frac{\omega
t}{2}\right) + \log \frac{\omega^2}{2\Lambda}
\end{equation}
respectively. And the scalar field evolves according to the following law
\begin{equation}
\phi = -\frac{k}{2} \log (\sinh \omega t)+\frac{m}{\omega}
\log \left(\tanh \frac{\omega t}{2}\right).
\end{equation}

It is interesting to see from this solution that the anisotropy is contributed
by both the scalar field (\(m\neq 0\)) and the transversal degree of the
gravitational field (\(a\neq 0\)). This can be seen as well from the Eq.~(23)
where the contribution to the anisotropy  due to the transversal degree of the
gravitational field and that due to the scalar field enter symmetrically.

These solutions decelerate for the times \(t<t_c\) where \(t_c\) is given by
\begin{equation}
t_c=\frac{1}{\omega} \cosh^{-1} \left( \frac{\frac{11}{6}m^2+\frac{3}{4}a^2}
{m\sqrt{2m^2+a^2}}\right)
\end{equation}
and then, after this time the solutions inflate.

\section{The Asymptotic Late-Time Behaviour of the Generic Solutions}

\subsection{The Numerical Analysis}

We now turn to study the asymptotic behaviour of the generic solutions of the
Eq.~(23). From the previous Section we have seen that  two different types of
late-times behaviour occur in exact solutions
\begin{equation}
G\sim t^N\,,
\end{equation}
and
\begin{equation}
G\sim e^{Nt}\, .
\end{equation}

These two different asymptotic behaviours correspond to FRW (Kasner if \(N=1\))
and to anisotropic models respectively. Technically we have found it very
difficult to study the integral (28) analytically. In most of the cases one can
not integrate the Eq.~(28) and further integrals in terms of  elementary
functions. We therefore have been forced to use numerical methods.

To ascertain as to whether the asymptotic behaviour described by the Eqs.~(44)
and (45) occurs in other solutions of the Eq.~(23) we have monitored the values
of the following functions while integrating the Eq.~(23) numerically
\cite{11}:
\begin{equation}
n_1(t)=\frac{\dot G^2}{\dot G^2-G\ddot G}\, ,
\end{equation}
together with
\begin{equation}
n_2(t)=\frac{\ddot G}{\dot G}\, .
\end{equation}

It is obvious that if a solution of the Eq.~(23) asymptotically follows the
power law given by the Eq.~(44) it is necessary and sufficient that the
function \(n_1\) defined by the Eq.~(46) tends to a constant as the time
increases. One could use a different function, say \(n(t)=t\dot G/G\), but our
numerical experiments have shown that the final answer is the same. The
quantity in (46) was finally retained because many test cases gave the answer
somewhat faster. Similarly the asymptotic behaviour given by the Eq.~(45) is
monitored by the test function \(n_2\) (or \(\dot G/G\) which finally gives
the same answer).

We have integrated numerically the Eq.~(23) for different values of the
constants \(k, m\) and \(a\). In each case different initial conditions for
\(G, \dot G\) and \(\ddot G\) were chosen, though always remaining positive
for physical reasons. Negative \(G\) would imply a change in the signature,
negative \(\ddot G\) would imply a negative cosmological constant via the
Eq.~(12) and finally the positivity of \(\dot G\) accounts for initial
expansion rather than contraction.

Our numerical results depict a clear scenario for the asymptotic behaviour
of the Eq.~(23). We have found that for every numerical solution the
constructed test function \(n_1\) tends towards a constant value which
depends only on the slope of the scalar field potential defined by the
constant \(k\). This behaviour and the value of the exponent \(N\) is
absolutely independent of the level of the anisotropy introduced either via
the scalar field, related to the parameter \(m\) or purely geometrical
anisotropy related with the parameter \(a\). For \(k^2<6\) the generic
solution behaves as
\begin{equation}
G\sim t^{\frac{4}{k^2-2}},
\end{equation}
as \(t\) goes to infinity which is an asymptotic behaviour of the isotropic
FRW model.

For \(k^2>6\) the following set of solutions of Eq.~(23)
\begin{equation}
G=Ct+D,
\end{equation}
which corresponds to an asymptotic behaviour of the vacuum Kasner solutions or
to the solutions with a minimally coupled massless scalar field, is an
attractor of the Eq.~(23) and describes the asymptotic behaviour of its generic
solution. It is worthwhile to mention that the status of expressions (48) and
(49) is different: while (48) is only an asymptotic solution, (49) is an exact
solution for an arbitrary set of constants \(K, M\) and \(B\).

We have also found that the asymptotic behaviour described by the Eq.~(45)
happens only in the case \(k^2=2\) and is structurally unstable, for any small
deviation in the parameter \(k\) was changing the asymptotic behaviour of the
solution to that described by the Eq.~(48).

\subsection{The Qualitative Analysis}

The outcome of our numerical calculations can be made plausible by a
qualitative
analysis of the Eq.~(23), which in terms of the new variables
\begin{equation}
x=\log G, \qquad y=\dot G,
\end{equation}
reduces to the following autonomous equation
\begin{equation}
y''\, +\, (K-1) y'=M^2 \,\frac{y'}{y^2}.
\end{equation}
In the particular case in which \(M=0\), the resulting linear equation
\begin{equation}
y''\, +\, (K-1) y'=0
\end{equation}
has the following general solution:
\begin{equation}
y\, = \, C \, + \, E e^{(1-K)x}.
\end{equation}

If \(K>1\) (i.e., if \(k^2>6\) ), and we assume that \(G\), and thus \(x\),
increases with time, the asymptotic behaviour of the general solution is
\(y\sim C\), which corresponds to the set of solutions given by the Eq.~(49).

In the non-linear case, \(M\neq 0\), one could expect that the asymptotic
behaviour is still the same because if \(y\rightarrow C \)
the right hand side of the Eq.~(51) goes to 0. Indeed, our numerical
experiments show that this is the case. This may be seen as well by
direct integration of the Eq.~(51), as depicted in the Fig.~1 for
\(K=2\) and \(M=0.1\). It is clearly seen that after some time the solutions of
the
Eq.~(51) become to a high precision straight lines with  a slope of  -1 (i.e.,
solutions
of the linear equation (52)) and, finally, approach the attracting line
\(y'=0\).

Though unchanging the asymptotic behaviour, the right
hand side of the Eq.~(51) is, of course, important for small \(y\).
To stress, it we have drawn as dashed lines some solutions of
the linear equation (52) corresponding to the same initial
conditions.

As for the assumption on \(G\) being an increasing function
of \(t\), we see in the same figure that solutions
corresponding to positive initial conditions (as required by
physical reasons) never reach negative values for \(y=\dot
G\), for they never cross the line \(y'=0\). This behaviour
was observed in all numerical integrations.

The cases corresponding to \(k^2<6 \,(-1/2<K<1)\) can be
discussed in  a very similar way by using
\begin{equation}
x\,=\,\log G, \qquad y\,=\,\dot G e^{(K-1)x},
\end{equation}
which leads to the following equation
\begin{equation}
y''\,+\, (1-K)\, y'\, = \, M^2\: \frac{y'+(1-K)y}{y^2}\,
e^{2(K-1)x}.
\end{equation}
Again, the particular case in which \(M=0\) is trivial and
its general solution, \(y=E+Fe^{(K-1)x}\), will approach
\(y=E\), which corresponds to (48). If \(M\neq 0\), \(y=D\)
is no longer a solution of the Eq.~(55) for a finite \(x\), but it can
still represent the asymptotic behaviour of the typical
solution, because the right hand side of the Eq.~(55) vanishes when
\(y\rightarrow E\) and \(x\rightarrow \infty\). We have
checked by numerical integration of both the Eqs.~(48) and (55) that this
is indeed the case.

\section{Conclusions}

We have presented here a general exact solution of the Eisntein field equations
for the anisotropic Bianchi type I universes filled with an
exponential-potential scalar field. Some of the representative cases and their
behaviour were explicitely considered in the Section IV. We must stress that by
considering and studying exact analytic examples one gets a good idea as to the
behaviour of the general model. Exact solutions, in our view, are indespensable
and must be considered, if possible, before any numerical analysis is
undertaken to be sure that the results based on numerical ``experiments'' are
of any significance.

Looking at the exact examples we have found three different typical late-time
behaviours of the models:
\newline i) The Kasner-like behaviour which is characteristic to the vacuum and
massless scalar field models.
\newline ii) The FRW-like behaviour, and
\newline iii) the limiting ``coasting'' type behaviour.

Integrating numerically the general solution we have seen that the limiting
\(G\sim e^{Nt}\) behaviour is structurally unstable. This in fact is very
interesting, for, exactly this type of behaviour as we shall see in paper II,
is generic for the models with one-dimensional inhomogeneity. So, while in such
simple models as Bianchi I the instability of these solutions do not cause any
reason to worry about the isotropization, in more complicated models their
stability causes problems.

We have seen that when the constant \(k^2\), defining the slope of the
potential is less than 6, the generic late-time behaviour is that of an
isotropic FRW model. In such situations the scalar field acts similarly to a
positive cosmological constant.

For \(k^2>6\) the late-time behaviour is that of a Kasner-type universe or,
which is the same, of the model filled with massless minimally coupled
scalar field. As long as \(k^2>6\) one has no reason to believe that the model
will isotropize.

To see whether the models inflate one may look at the sign of the deceleration
parameter \(q\). After some algebra and using the first integral given by the
Eq.~(27) the sign of the deceleration parameter \(q\) is given by
\begin{equation}
\frac{3}{2}\,K\dot G^2-\left( km+\frac{B}{2}\right)\,\dot G+\frac{1}{6}\,
(km-B)^2+\frac{3}{2}\, M.
\end{equation}
It is easy to see from this expression that as long as the anisotropy
parameters \(M\) and \(B\) are switched off (\(m=a=0\)), the inflation of the
solutions depends only on slope of the potential given by \(k\). Nevertheless,
if the
anisotropy is present the inflation is not only driven by the parameter \(k\)
but depends as well on the rates of anisotropy. Studying the behaviour of the
Eq.~(50) one finds out that solutions generically inflate for \(k^2<2\)
confirming previous results. For \(k^2>2\) most of the solutions do not
inflate,  yet depending on the rate of the anisotropy one may find inflating
solutions.
\\[1cm]
{\bf Acknowledgments}

Two of us (A.F. and J.I.) are grateful to Prof. M.A.H. MacCallum and to the
members of the QMW College Relativity Group for their hospitality and
stimulating discussions. This work was supported by the CICYT grant PS90-0093.

\newpage
\centerline{\bf Figure Caption}
\vspace{2cm}
Fig. 1~~~Phase-space of the Eq.~(51) for \(K=2\) and \(M=0.1\). The solutions
correspond to initial conditions in the form \(y_0=0.01, y'_0=0.01+0.1n\),
with \(n=0\ldots 12\). The dashed lines represent the solutions of the linear
case
(52) for the same initial conditions for \(n=1\ldots 7\).


\begin{thebibliography}{10}

\bibitem[1]{1} A.~Feinstein and J.~Ib\'a\~nez, Class. Quantum Grav. {\bf 10}
(1993) 93.
\bibitem[2]{2} A.~Feinstein and J.~Ib\'a\~nez, {\em Exact Inhomogeneous Scalar
Field Universes} submitted.
\bibitem[3]{3} R.M.~Wald, Phys. Rev. D {\bf 28} (1983) 2118.
\bibitem[4]{4} C.W.~Misner, Ap. J. {\bf 151} (1968) 431.
\bibitem[5]{5} K.A.~Olive, Phys. Rep. {\bf 190} (1990) 307.
\bibitem[6]{6} J.J.~Halliwell, Phys. Lett. B {\bf 185} (1987) 341.
\bibitem[7]{7} A.B.~Burd and J.D.~Barrow, Nucl. Phys. B {\bf 308} (1988) 929.
\newline A.B.~Burd, {\em General Relativity Scalar Fields and Cosmic Strings},
D. Phil. Thesis, University of Sussex (1987).
\newline M.S.~Turner and L.M.~Widrow, Phys. Rev. Lett. {\bf 57} (1986) 2237.
\newline Y.~Kitada and K.~Maeda, Phys. Rev. D {\bf 45} (1992) 1416.
\newline V.~M\"{u}ller, H.J.~Schmidt and A.A.~Starobinsky, Class. Quantum Grav.
{\bf 7} (1990) 1163.
\newline Y.~Kitada and K.~Maeda, Class. Quantum Grav. {\bf 10} (1993) 703.
\bibitem[8]{8} G.F.R.~Ellis, in {\em Cargese Lectures in Physics} vol.6, Ed.
E.~Schatzmann, Gordon and Breach (1973).
\bibitem[9]{9} C.B.~Collins and S.W.~Hawking, Ap. J. {\bf 180} (1973) 317.
\bibitem[10]{10} G.F.R.~Ellis and M.S.~Madsen, Class. Quantum Grav. {\bf 8}
(1991) 667.
\bibitem[11]{11} The numerical integrations were performed by means of {\it ODE
Workbench} (J.M.Aguirregabiria, {\it ODE Workbench}, Physics Academy Software
(AIP), in press (1993)). The quality of the numerical results was tested by
using different integration codes, ranging from the very stable embedded
Runge-Kutta code of eighth order due to Dormand and Prince to very fast
extrapolation routines. All the codes have adaptive step size control and we
checked that smaller tolerances did not change the results. Double precision
was used in all calculations and different exact cases were used to test our
numerical work.

\end{thebibliography}
 \end{document}